\documentclass[a4paper,twocolumn]{autart}

\usepackage{tikz}
\usepackage{verbatim}
\usetikzlibrary{arrows}

\usetikzlibrary{%
    decorations.pathreplacing,%
    decorations.pathmorphing%
}

\usetikzlibrary{decorations.pathmorphing} 
\usetikzlibrary{matrix} 
\usetikzlibrary{arrows} 
\usetikzlibrary{calc} 
\usetikzlibrary{shapes.geometric, arrows}
\tikzstyle{block} = [draw,rectangle,thick,minimum height=2em,minimum width=2em]
\tikzstyle{sum} = [draw,circle,inner sep=0mm,minimum size=2mm]
\tikzstyle{connector} = [->,thick]
\tikzstyle{line} = [thick]
\tikzstyle{branch} = [circle,inner sep=0pt,minimum size=1mm,fill=black,draw=black]
\tikzstyle{guide} = []
\tikzstyle{snakeline} = [connector, decorate, decoration={pre length=0.2cm,
                         post length=0.2cm, snake, amplitude=.4mm,
                         segment length=2mm},thick, magenta, ->]

\usetikzlibrary{decorations.markings}

\renewcommand{\vec}[1]{\ensuremath{\boldsymbol{#1}}} 

\usepackage{amsmath,amssymb,enumerate}
\usepackage{graphicx}
\usepackage{color}
\usepackage{indentfirst} 

\usepackage{booktabs}
\usepackage{multirow}
\usepackage[normalem]{ulem}
\usepackage{soul} 
\soulregister\cite7
\soulregister\ref7
\soulregister\pageref7

\newtheorem{definition}{Definition}

\newtheorem{theorem}{Theorem}
\newtheorem{remark}{Remark}

\newtheorem{corollary}{Corollary}

\usepackage{graphicx}
\usepackage{caption}
\usepackage{subfig}

\begin{document}
\begin{frontmatter}

\title{Event-triggered leader-following tracking control for
  multivariable multi-agent systems\thanksref{footnoteinfo}}

\thanks[footnoteinfo]{This work was supported by the
    Australian Research Council under the Discovery Projects funding scheme
    (project DP120102152). Accepted for publication in Automatica on 
    March 14, 2016.}

\date{Accepted for publication in Automatica on 
    March 14, 2016.}

\author[ADFA]{Yi Cheng}\ead{y.cheng@adfa.edu.au} \and
\author[ADFA]{V.~Ugrinovskii}\ead{v.ugrinovskii@gmail.com}

\address[ADFA]{School of Engineering and Information Technology,
University of New South Wales at the Australian Defence Force Academy,
Canberra, ACT 2600, Australia.}


\begin{abstract}

The paper considers event-triggered leader-follower tracking control for
multi-agent systems with general linear dynamics. For both undirected and
directed follower graphs, we propose event triggering rules which guarantee
bounded tracking
errors. With these rules, we also prove that the systems do not exhibit Zeno
behavior, and the bounds on the tracking errors can be tuned to a desired
 small value. We also show that the combinational
state required for the proposed event triggering conditions can be
continuously generated from discrete communications between the neighboring
agents occurring at event times. The efficacy of the proposed methods is
discussed using a simulation example.

\end{abstract}

\end{frontmatter}

\section{Introduction}
Cooperative control of multi-agent systems has received increasing
attention in the past decade, see~\cite{Ren2008} and references therein. However, many control techniques developed so
far rely on continuous
communication between agents and their neighbors. 
This limits practicality of these techniques.

To address this concern, several approaches have been proposed in recent years.
One approach is to apply sampled control \cite{Xie2009}. However in sampled
data control schemes control action updates continue periodically with the
same frequency even after the system has reached the control goal with
sufficient accuracy and no longer requires intervention from the
controller. Efforts to overcome this 
shortcoming have led to the idea of triggered control.
Self-triggered control strategies \cite{Mazo2010,Heemels2012,Dimarogonas2012}
employ a triggering mechanism to proactively predict the next time for
updating the control input ahead of time, using the current
measurements. On the other hand, event-triggered controllers
\cite{Tabuada2007,Lunze2010,Dimarogonas2009,Wang2011,Dimarogonas2012,Heemels2013}
trigger control input updates by reacting to excessive deviations of the
decision variable from an acceptable value, i.e., when a continuously
monitored triggering condition is violated. This latter approach is the main focus in this paper.

The development of event-triggered
controllers remains challenging, because the agents in a multi-agent
system do not have access to the complete system state
information required to make decisions about control input updates. 
To prove the concept of event-triggering, the early work was still
assuming continuous communication between 
the neighboring agents \cite{Dimarogonas2009,Dimarogonas2012}. To
circumvent this limitation, several
approaches 
have been proposed, e.g., see \cite{Seyboth2013,Fan2013,Meng2013,Zhu2014,Garcia2014,Liuzza2013,Adaldo2014}.
For instance, different from \cite{Dimarogonas2009,Dimarogonas2012} where
state-dependent event triggering conditions were used, \cite{Seyboth2013}
proposed an event-triggered control strategy using
a time-dependent triggering function which did not require neighbors'
information. In \cite{Fan2013},
a state-dependent event triggering condition was employed, 
complemented by an iterative algorithm to estimate the next
triggering time, so that continuous communications between neighboring
agents were no longer needed.
In~\cite{Meng2013}, sampled-data event detection has been used.
It must be noted that these results as well as many other results in
this area
were developed for multi-agent systems with single or double
integrator dynamics. Most recently, similar results have been developed for multi-agent systems with general dynamics
\cite{Zhu2014,Garcia2014} and nonlinear dynamics \cite{Liuzza2013,Adaldo2014}.

All the papers mentioned above considered the event-triggered
control problem for leaderless systems. The leader-following 
control is one
of the important problems in cooperative control of multi-agent systems
\cite{Jadbabaie2003,Ren2007,Hong2006,Ren2008}, and
the interest in event-based solutions to this problem is growing \cite{Hu2011,Zhang2012,Li2015,Hu2015}. General multidimensional leader following problems
still remain technically challenging, and the development is often restricted to
the study of single or double integrator
dynamics~\cite{Hu2011,Zhang2012,Li2015,Hu2015}. Zeno
behavior presents another
challenge, and is not always excluded \cite{Hu2011,Zhang2012}. Excluding Zeno
behavior is an important requirement on control protocols
since excessively frequent communications reduce the advantages of using the
event-triggered control.

In this paper, we also consider the event-triggered leader-following control problem for multi-agent systems. Unlike \cite{Hu2011,Zhang2012,Li2015,Hu2015}, the class of
systems considered allows for general linear dynamics. Also, the leader can be marginally stable or even unstable. For both undirected
and directed system interconnections, we propose sufficient
conditions for the design of controllers which guarantee that the leader
tracking errors are contained within certain bounds; these bounds can be
optimized by
tuning the parameters of the design procedure.
We also show that with the proposed event-triggered control protocols, the system does not exhibit Zeno behavior. These results are
the main contribution of the paper.

Its another contribution is the event-triggered control
protocols that do not require the neighboring agents to communicate
continuously. Instead, the combinational state to be used in the event
triggering condition is generated continuously within
the controllers, by integrating
the information obtained from the neighbors during their
communication events. The idea 
is inspired by \cite{Fan2013}, however,
the procedure in~\cite{Fan2013} developed for single integrator systems
cannot be applied to multi-agent systems with general linear dynamics
considered here, since in our case dynamics of the measurement error depend
explicitly on the combinational state.
Also different from 
\cite{Fan2013}, the proposed algorithm involves
one-way communications between the neighboring agents. The combinational state is computed continuously by each agent and
is broadcast to its neighbors \emph{only at the time when the communication
  event is triggered} at this node and \emph{only in one direction}. The
neighbors then use this information for their own computation, and do not
send additional requests to measure the combinational state. This is an important advantage of our protocol compared with event-triggered control strategies proposed in~\cite{Fan2013,Hu2011,Li2015,Hu2015,Zhu2015}. In these references, when an event is triggered at one agent, it must request its neighbors for additional information to update the control signals. Owing to this, our scheme is applicable to systems with a directed graph which only involves one way communications.

In comparison with the recent work on event-triggered control for general
linear systems~\cite{Zhu2014,Zhu2015,Garcia2014,Liu2013},
the main distinction of our method is computing the combinational state
directly using the neighbors' information. This allowed us to
avoid additional sampling when checking event triggering conditions,
cf.~\cite{Zhu2014,Zhu2015}. In contrast in \cite{Garcia2014}, to avoid continuous
transmission of information, each agent was equipped with 
models of itself and its neighbors. In~\cite{Liu2013},
estimators were embedded into each node  to
enable the agents to estimate their neighbors' states. Both approaches make
the controller rather complex, compared with our controller
which does not require additional  models or
estimators. The leader-follower context and the treatment of both directed and undirected versions of the problem are other distinctions.

The paper is organized as follows. Section~\ref{problem formulation}
includes the problem formulation and preliminaries. The main results
are given in Sections~\ref{main.2} and~\ref{main.1}. In
Section~\ref{main.2} we consider the case when the system of followers is
connected over a directed graph. Although these results are applicable to
systems connected over an undirected graph as
well, the symmetry of the graph Laplacian makes it possible to derive an
alternative control design scheme in Section~\ref{main.1}. In
Section~\ref{algorithm}, the generation of the combinational
state is discussed. Section~\ref{example} provides an illustrative
example. The conclusions are given in Section~\ref{conclusion}.

Throughout the paper, $\Re^n$ and $\Re^{n\times m}$ are a real Euclidean $n$-dimensional vector
space and a space of real $n\times m$ matrices. $\otimes$ denotes the Kronnecker product of two matrices.
$\lambda_{\max}(\cdot)$ and $\lambda_{\min}(\cdot)$ will denote the largest
and the smallest eigenvalues of a real symmetric matrix. For $q\in \Re^n$, $\mathrm{diag}\{q\}$ denotes the diagonal matrix with the
entries of $q$ as its diagonal elements. $I_N$ is the $N\times N$ identity matrix. When the dimension is
clear from the context, the subscript $N$ will be suppressed.

\section{Problem formulation and preliminaries}
\label{problem formulation}

\subsection{Communication graph}

Consider a communication graph $\mathcal {\bar G}=(\mathcal {\bar V}, \mathcal {\bar E},
\mathcal {\bar A})$, where $\mathcal {\bar V}= \{0, \ldots, N\}$ is a finite nonempty node set, $\mathcal
{\bar E} \subseteq \mathcal {\bar V}\times \mathcal {\bar V}$ is an edge set of pairs of nodes, and $\mathcal{\bar A}$ is an adjacency matrix. Without loss of generality, node $0$ will be assigned
to represent the leader, while the nodes from the set $\mathcal {V}= \{1,
\ldots, N\}$ will represent the followers.

The (in general, directed) subgraph $\mathcal {G}=(\mathcal {V}, \mathcal {E},
\mathcal {A})$ obtained from $\mathcal {\bar G}$ by removing the leader
node and the corresponding edges describes communications
between the followers; the edge set $\mathcal{E} \subseteq \mathcal {V}\times \mathcal {V}$ represents the communication
links between them, with the ordered pair $(j,
i)\in\mathcal {E}$ indicating that node $i$ obtains information from node
$j$; in this case $j$ is the neighbor of $i$. The set of
neighbors of node $i$
in the graph $\mathcal{G}$ is denoted as $N_i=\{j|(j, i)\in \mathcal
{E}\}$. Following the standard convention, we assume that $\mathcal {G}$
does not have self-loops or repeated
edges. The adjacency matrix $\mathcal {A}=[a_{ij}]\in \Re^{N\times N}$
of $\mathcal {G}$ is defined as $a_{ij}=1$ if
$(j, i)\in \mathcal {E}$, and $a_{ij}=0$ otherwise. 
Let $d_i = \sum_{j=1}^{N}a_{ij}$ be the in-degree
of node $i\in\mathcal{V}$ and
$\mathcal{D}=\mathrm{diag}\{d_1,\ldots,d_N\}\in \Re^{N\times N}$. Then
$\mathcal {L} =\mathcal {D} - \mathcal {A}$ is the Laplacian matrix of
the graph $\mathcal {G}$, it is symmetric when $\mathcal {G}$ is undirected. 

We assume throughout the paper that the
leader is observed by a subset of followers. If the leader is observed by
follower $i$, then the directed edge $(0,i)$ is included in
$\bar{\mathcal{E}}$ and is assigned with the weighting $g_i=1$, otherwise we
 let $g_i=0$. We refer to node $i$ with
$g_i\neq 0$ as a pinned node. Let $G=\mathrm{diag}\{g_1, \ldots, g_N\}\in \Re
^{N\times N}$. The system is assumed to have at least one follower which can observe the leader, hence $G \neq 0$.

In addition, we assume the graph $\mathcal{G}$ contains a
spanning tree rooted at a pinned node $i_r$, i.e., $g_{i_r}>0$. Then, $-(\mathcal{L}+G)$ is a Metzler matrix. According to~\cite{Hu2007}, the matrix $-(\mathcal{L}+G)$ is Hurwitz stable\footnote{These properties of the matrix $\mathcal{L}+G$ can be guaranteed under weaker assumptions on the graph
$\mathcal{G}$~\cite{Hu2007}.}, which implies that $-(\mathcal{L}+G)$ is diagonally stable~\cite{Kaszkurewicz2000}. That is, there exists a positive definite diagonal matrix $\Theta= \mathrm{diag}\{ \vartheta_1, \ldots, \vartheta_N\}$ such that $H=\Theta^{-1}(\mathcal{L}+G) + (\mathcal{L}+G)'\Theta^{-1} >0$. We will also use the following notation: $\alpha =\frac{1}{2}\lambda_{\min}(H)$, $\vartheta_{\min}= \min \limits_i(\vartheta_i)$, $\underline{\vartheta}= \min \limits_i(\vartheta_i^{-1})$, $P=\Theta^{-1}(\mathcal{L}+G)(\mathcal{L}+G)'\Theta^{-1}$ and $F=(\mathcal{L}+G)'(\mathcal{L}+G)$.


\subsection{Problem formulation}
 Consider a multi-agent system consisting of a leader agent and $N$ follower agents. Dynamics of the $i$th follower are described by the equation
\begin{align} \label{agents dymamic}
 \dot{x}_i=Ax_i + Bu_i,
\end{align}
where $x_i\in \Re^n$ is the state, $u_i\in \Re^p$ is the control input. Also, the dynamics of the leader agent are given by
\begin{align} \label{leader dymamic}
 \dot{x}_0=Ax_0.
\end{align}
Note that the matrix $A$ is not assumed to be Hurwitz, it can be marginally stable or even unstable.

We wish to find a distributed event-triggered
control law for each follower to be able to track the leader. For each agent $i$, introduce a
combinational state $z_i(t)$,
\begin{align}
\label{define z_i}
z_i(t)=\sum\limits_{j\in N_i} \big( x_j(t)-  x_i(t) \big) + g_i\big(x_0(t) -  x_i(t)\big).
\end{align}
We seek to develop a control scheme where agent $i$ updates its control
input at event times, which are denoted by $t_0^i,t_1^i, \ldots$, based on
samples $z_i(t_k^i)$ of its combinational state. The value of the
combinational state is held constant between updates,
thus giving rise to the measurement signal
$\hat z_i(t)=z_i(t_k^i), ~~t \in [t_k^i,t_{k+1}^i)$.
Based on this model, consider the following control law
\begin{align} \label{controller}
 u_i(t)=-K \hat z_i(t), ~~t \in [t_k^i,t_{k+1}^i),
\end{align}
where $K \in \Re^{p\times n}$ is a feedback gain matrix to be defined later.
The problem in this paper is to find a control law
(\ref{controller}) and an event triggering strategy which achieve the
following leader-following property
\begin{align}
\label{defined tracking consensus}
\limsup_{t\rightarrow \infty}{\textstyle\sum_{i=1}^{N}}   \|x_0(t) - x_i(t)\|^2 \le \Delta,
\end{align}
where $\Delta$ is a given positive constant. Furthermore, the closed loop
dynamics of the followers must not exhibit Zeno behavior with the
proposed event triggering rule.

\begin{definition}
  We say that the leader-follower system (\ref{agents dymamic}),
  (\ref{leader dymamic}) with a control law
   (\ref{controller}) does not exhibit Zeno behavior if over
  any finite time period there are only a finite number of communication
  events between the follower systems, i.e., for every agent $i$ the
  sequence of event times $t_k^i$ has the property $\inf_{k}
  (t_{k+1}^i - t_k^i ) >0$.
\end{definition}

\section{Event-triggered leader-following control under a directed graph $\mathcal{G}$}
\label{main.2}

In this section, we propose an event triggering rule and a leader-following
tracking control for multi-agent systems where the followers are connected
over a directed graph. Our result will involve
certain symmetric positive definite matrices $R$, $Q$, and $Y$ related
through the following Riccati inequality
\begin{align}
\label{ARI2}
YA + A'Y - 2 \vartheta_{\min} YB R^{-1}B'Y +  Q \le 0,
\end{align}
 and constants $\omega >0$ and $\mu_i >0$ chosen so that $\rho_i = \alpha_1
-\mu_i -  \frac{\omega}{\alpha} \alpha_2 >0$, where $\alpha_1= \frac{\lambda_{\min}(Q)}{\lambda_{\max}(Y)}$ and
$\alpha_2 = \frac{\lambda_{\max}(P)\lambda^2_{\max}(YBR^{-1}B'Y)}{\underline{\vartheta} \lambda_{\min}(Y)}$. Let
$\rho_{\min}= \min \limits_i \rho_i$ and
select $\nu_i>0$, $\sigma_i\in(0,\rho_{\min})$ and $\gamma>0$.
Introduce the combinational state measurement error for agent $i$
\begin{align}
\label{measure error}
  s_i(t)= \hat z_i(t) - z_i(t).
\end{align}

\begin{theorem}\label{T2}
Given $R=R'>0$, $Q=Q'> 0$, suppose
there exists  $ Y= Y'> 0$ such that (\ref{ARI2}) holds.
Then under the control law (\ref{controller}) with $K=- \frac{1}{\alpha}R^{-1}B'Y$, the system (\ref{agents dymamic}),
 (\ref{leader dymamic}) achieves the leader-follower tracking property of the form
(\ref{defined tracking consensus}) with $\Delta=\frac{N \gamma}{ \underline{\vartheta} \lambda_{\min}(Y) \lambda_{\min}(F)\rho_{\min}}$,
if the
communication events are triggered at
\begin{align}
\label{event condition TH2}
\hspace{-.3cm}t_k^i=\inf \Big \{ & t>t_{k-1}^i \colon
\nonumber \\
&\|s_i\|^2 \ge {\alpha \omega }(\mu_i \vartheta_i^{-1} z_i' Y
z_i + \nu_i e^{-\sigma_i t} + \gamma) \Big \}.
\end{align}
In addition, the system does not exhibit Zeno behavior.
\end{theorem}

\begin{remark}\label{LMI.equiv} \em
The Riccati inequality (\ref{ARI2}) is similar to the Riccati inequality
employed in \cite{Garcia2014}. However, \cite{Garcia2014} considers an undirected topology, and the design uses the second
smallest eigenvalue of $\mathcal{L}$. In contrast, (\ref{ARI2}) uses
$\vartheta_{\min}$ associated with the directed
graph $\mathcal{G}$.
Also, the inequality
(\ref{ARI2}) is equivalent to the following LMI in $Y^{-1}$, which can be
solved using the existing LMI solvers,
\begin{align}
\label{LMI.directed}
\left[\begin{array}{cc}
   AY^{-1} +Y^{-1}A' - 2 \vartheta_{\min} BR^{-1}B     &   Y^{-1}\\
    Y^{-1}                &     -Q^{-1}
\end{array}\right] \le 0.
\end{align}
\end{remark}

\begin{remark} \em
We note that the event
triggering condition (\ref{event condition TH2}) involves monitoring of the
combinational state $z_i(t)$, hence the means for
generating $z_i(t)$ continuously are needed to implement it. A
computational algorithm will be
introduced later to generate the combinational state using event-triggered
communications.
\end{remark}

\emph{Proof of Theorem~\ref{T2}:}
We first prove that $\|z_i(t)\|$ are bounded. This fact will then be used
to prove that under the proposed control law the system does not exhibit
Zeno behavior. Also, the property (\ref{defined tracking consensus}) will
be proved after Zeno behavior is excluded.

Define the tracking error $\varepsilon_i(t)= x_0(t) - x_i(t)$ at
node $i$. It follows from (\ref{measure error}) and (\ref{controller}) that
\begin{align} \label{error dymamic with norm}
 \dot{\varepsilon}_i(t) =& A\varepsilon_i(t) + BK z_i(t) + BK s_i(t) .
\end{align}
Let the Lyapunov function candidate for the system
comprised of the systems (\ref{error dymamic with norm}) be $
V(\varepsilon)=  z'(\Theta^{-1} \otimes Y)z$,
where $z=((\mathcal {L} +G)\otimes I_n) \varepsilon$, and
$\varepsilon=[\varepsilon_1'~\ldots~\varepsilon_N']'$. Then
\begin{align}
\label{lyapunov equation2.1}
&\frac{d V(\varepsilon)}{dt} = 2 z' \big( (\Theta^{-1}(\mathcal {L} +G) \otimes YBK)z \nonumber\\
&+ (\Theta^{-1}\otimes YA)z + (\Theta^{-1} (\mathcal {L} +G) \otimes YBK)s \big).
\end{align}
Since $K= - \frac{1}{\alpha} R^{-1}B'Y$, the following inequality holds
\begin{eqnarray}
\label{directed_trick}
 \lefteqn{2 z' ((\Theta^{-1}(\mathcal {L} +G))\otimes (YBK))z} && \nonumber\\
 &&=  -z'\Big( H \otimes (\frac{1}{\alpha} YBR^{-1}B'Y)\Big)z \nonumber\\
 && \le -2 \alpha z'\Big( I_N \otimes (\frac{1}{\alpha} YBR^{-1}B'Y)\Big)z.
\end{eqnarray}
Using  (\ref{ARI2}),
it follows from (\ref{lyapunov equation2.1}) and (\ref{directed_trick}) that
\begin{align}
\label{lyapunov equation2.21}
\frac{d V(\varepsilon)}{dt} \le &- z' \big(\Theta^{-1} \otimes Q \big)z   + \frac{1}{\alpha \omega}s's \nonumber\\
&+  \frac{\omega}{\alpha}z' \big(P \otimes (YBR^{-1}B'Y)^2\big)z.
\end{align}
Since the triggering condition (\ref{event condition TH2}) enforces the property
\begin{align}
\label{lyapunov equation2.23}
(\alpha \omega)^{-1}  \|s_i\|^2 \le \mu_i \vartheta_i^{-1} z_i' Y z_i + \nu_i e^{-\sigma_i t} + \gamma
\end{align}
on every interval $[t_k^i,~t_{k+1}^i)$, then it follows from (\ref{lyapunov equation2.21}) that
\begin{align}
\label{VO.dVdz}
\frac{d V(\varepsilon)}{dt} &\le -\sum_{i=1}^{N} \rho_i \vartheta_i^{-1} z_i' Y z_i + \sum_{i=1}^{N} \nu_i e^{-\sigma_i t} + N\gamma \nonumber\\
& \le - \rho_{\min} V(\varepsilon)  +\sum_{i=1}^{N} \nu_i e^{-\sigma_i t} + N\gamma.
\end{align}
Thus, we have
\begin{eqnarray}
V(\varepsilon) &\le  e^{- \rho_{\min} t} \Big(V(\varepsilon(0))- \sum_{i=1}^{N} \frac{\nu_i}{\rho_{\min} - \sigma_i} - \frac{N \gamma}{\rho_{\min}}\Big)  \nonumber\\
& + \sum_{i=1}^{N} e^{- \sigma_i t}\frac{\nu_i}{\rho_{\min}-\sigma_i} +
\frac{N \gamma}{\rho_{\min}}
\label{VO.dVdz_2.a} \\
&\le V(\varepsilon(0)) + \sum_{i=1}^{N}  \frac{\nu_i}{\rho_{\min}-\sigma_i} + \frac{N \gamma}{\rho_{\min}} = \kappa,
\label{VO.dVdz_2}
\end{eqnarray}
where the constant $\kappa$ depends on the initial conditions.
It then follows from (\ref{VO.dVdz_2}) that for all $t\ge 0$
\begin{align}
\label{lyap-consensus2}
\kappa \ge V(\varepsilon)= z'(\Theta^{-1} \otimes Y) z \ge \underline{\vartheta} \lambda_{\min}(Y) \sum_{i=1}^{N} \| z_i \|^2.
\end{align}
This implies that for all $i$, $\|z_i(t)\|$ is bounded,
\begin{align}
\label{bounded z_i}
 \| z_i(t) \| \le 
\sqrt{{\kappa}{ (\underline{\vartheta} \lambda_{\min}(Y)})^{-1}}=\bar \kappa.
 \end{align}

Next, we prove that the system does not exhibit Zeno behavior.
Suppose $t_1$, $t_2$ are two adjacent zero points of $s_i(t)$ on the
interval $[t_k^i, t_{k+1}^i)$, $t_k^i\le t_1<t_2< t_{k+1}^i$. Then
$\|s_i(t)\| > 0$ for all $t \in (t_1, t_2)\subseteq(t_k^i, t_{k+1}^i)$, and the
following inequality holds on the interval $(t_1, t_2)$
 \begin{align}
 \label{ds_dt_directed}
 \frac{d}{dt}\|s_i\| =  \frac{d}{dt}(s_i's_i)^{1/2} =  \frac{s_i' \dot s_i}{\| s_i \|} \le  \frac{\|s_i\| \|\dot s_i\|}{\| s_i \|} =  \|\dot s_i\|.
 \end{align}
Furthermore, note that on
the interval $[t_1, t_2)$
\begin{equation}
\label{dot s_i.1_directed}
 \dot s_i(t)=- \dot z_i(t),  \quad s_i(t_1^+)\footnote{As usual,
  $s(a^+)\triangleq\lim_{t\downarrow a}s(t)$, $s(b^-)\triangleq\lim_{t\uparrow b}s(t)$.}=0.
 \end{equation}
It follows from (\ref{dot s_i.1_directed}) that $\forall t \in [t_1, t_2)$
\begin{align}
\label{bound_dot s_i_directed}
  &\|\dot s_i(t)\| =  \big\| \sum\limits_{j\in N_i} \big( \dot x_j(t)-  \dot x_i(t) \big) + g_i \big( \dot x_0(t)-  \dot x_i(t) \big) \big\|  \nonumber \\
 &=  \Big\|  A z_i (t) + BK\Big ( \sum\limits_{j\in N_i} \big(z_i(t_k^i) -\hat z_j(t)\big)  + g_i z_i(t_k^i)  \Big)  \Big\|   \nonumber \\
 &\le \|A\| \|s_i (t)\| + M_k^{i},
\end{align}
where $M_k^{i}= \max \limits_{t \in [t_k^i, t_{k+1}^i)} \| A z_i(t_k^i) +
BK\Big ( \sum\limits_{j\in N_i} \big(z_i(t_k^i) -\hat z_j(t)\big)  + g_i
z_i(t_k^i)\Big)\|$. Hence, using (\ref{ds_dt_directed}) and (\ref{bound_dot
  s_i_directed}) we obtain
\begin{align}
\label{s_i_directed_t12}
\hspace{-2ex}  \|s_i\|\le  \frac{ M_k^{i}}{\|A\|} (e^{\|A\|(t-t_1)} - 1)
\le  \frac{ M_k^{i}}{\|A\|} (e^{\|A\|(t_{k+1}^i-t_k^i)} - 1)
\end{align}
for all $t\in (t_1, t_2)$.
Since $s_i(t_1^+)=0$, (\ref{s_i_directed_t12}) holds for all $t\in [t_1,
t_2)\subseteq [t_k^i, t_{k+1}^i)$. The expression on the right hand side of
(\ref{s_i_directed_t12}) is independent of $t$; hence the above reasoning
applies to all such intervals $[t_1, t_2)$. Hence,  (\ref{s_i_directed_t12}) holds for all $t \in [t_k^i, t_{k+1}^i)$. Thus,
from the definition of the event time $t_{k+1}^i$ in (\ref{event condition
  TH2}) and (\ref{s_i_directed_t12}) we obtain
\begin{align}
\label{lower bounded.directed}
\hspace{-2ex} \sqrt{\alpha \omega\gamma} \le \|s_i((t_{k+1}^i)^-)\| \le  \frac{ M_k^{i}}{\|A\|} (e^{\|A\|(t_{k+1}^i-t_k^i)} - 1).
\end{align}
According to (\ref{bounded z_i}),  for any $k$,
$M_k^{i} \le
\big(\| A \| +  (2d_i+ g_i) \| BK\|\big) \bar \kappa
  = \eta_i \bar \kappa \le \bar \eta  \bar \kappa$, where $\bar \eta =\max
  \limits_i \eta_i$.
Hence, it follows from (\ref{lower bounded.directed}) that
\begin{align}
\label{lower bounded_t_k_i.directed}
 t_{k+1}^i - t_k^i \ge  \frac{1}{\|A\|} \ln \Big(1+ \frac{\|A\|  \sqrt{\alpha \omega \gamma}}{\bar \eta  \bar \kappa} \Big).
\end{align}
Thus, the inter-event intervals are bounded from below uniformly in $k$,
that is, Zeno
behavior does not occur.

Since Zeno behavior has been ruled out, it follows from
(\ref{VO.dVdz_2.a}) and the rightmost inequality in (\ref{lyap-consensus2})
that for all $i$,
\begin{align}
\label{bounded_z_i2_infty}
\limsup  \limits_{t\to \infty} {\textstyle\sum_{i=1}^{N}} \| z_i(t)\|^2 \le  
{N \gamma}{ (\underline{\vartheta} \lambda_{\min}(Y) \rho_{\min})^{-1}}.
\end{align}
Since $z=((\mathcal {L} +G)\otimes I_n) \varepsilon$, this further implies
\begin{align}
\label{bounded_e_i.directed}
\limsup_{t\to \infty} {\textstyle\sum_{i=1}^{N}} \| \varepsilon_i(t)\|^2 \le  \frac{N \gamma}{ \underline{\vartheta} \lambda_{\min}(Y) \lambda_{\min}(F)\rho_{\min}}.
\end{align}
I.e.,~(\ref{defined tracking consensus}) holds. This concludes the proof.
\hfill$\Box$

According to (\ref{lower bounded_t_k_i.directed}) and
(\ref{bounded_e_i.directed}), the parameter $\gamma$ not only helps to
exclude Zeno behavior, but also determines  the upper bound of the
tracking errors. We now show that after a sufficiently large
time, the lower bound on the inter-event intervals becomes independent of
$\gamma$. More precisely,  the following statement holds.
\begin{corollary}
\label{coro1}
For any $\delta>0$, there exists a sufficiently large $t_\delta$ such that with the
control law and event triggering condition proposed in
Theorem~\ref{T2},
\begin{align}
\label{lower bounded_s_k.directed}
&\hspace{-3ex}\inf_{k\colon t_k^i>t_\delta}(t_{k+1}^i - t_k^i) \nonumber \\
& \ge  \frac{1}{\|A\|} \ln \Big(1
+ \frac{\|A\|  \sqrt{\alpha \omega \underline{\vartheta} \lambda_{\min}(Y) \rho_{\min}} }{\bar \eta (1+\delta) \sqrt{N} }  \Big)=\pi.
\end{align}
\end{corollary}

\emph{Proof:}
According to (\ref{bounded_z_i2_infty}), for
any $\delta>0$, there exists $t_{\delta}$ such that
$\|z_i(t)\|< (1+\delta)\sqrt{\frac{N \gamma}{ \underline{\vartheta} \lambda_{\min}(Y) \rho_{\min}}}\triangleq \varpi_1$
for all $i$ and $t>t_{\delta}$.
Therefore, for a sufficiently large $k$, $M_k^{i} \le
\big(\| A \| +  (2d_i + g_i) \| BK\|\big) \varpi_1
  \le \bar \eta \varpi_1$. Then (\ref{lower bounded_s_k.directed}) follows
  from (\ref{lower bounded.directed}).
\hfill$\Box$

\begin{remark} \em
\label{remark Th.directed}
From (\ref{bounded_e_i.directed}),  the upper bound on the
tracking error depends on the parameter $\gamma$ and the size of the
network $N$. Therefore, the tracking performance can be guaranteed even for
larger systems, if $\gamma$ is sufficiently small. On the other hand,
the lower bound on the inter-event times in (\ref{lower
  bounded_t_k_i.directed}) reduces if $\gamma$ is reduced. This means that a higher tracking precision
  can be achieved by reducing $\gamma$, but the communications may become
  more frequent. However,
Corollary~\ref{coro1} shows that when $\gamma$ is reduced, the frequency of
communication events may increase
only on an initial interval $[0,t_\delta]$, and after time
$t_\delta$ the minimum inter-event time $\pi$ is independent of $\gamma$.
\end{remark}

\begin{remark} \em
\label{remark event.paramters.directed}
Selecting the parameters for the event-triggering condition (\ref{event
  condition TH2}) involves the following steps:
(a) choose matrices $Q>0$ and $R>0$ and solve the Riccati
inequality~(\ref{ARI2}), or equivalently the LMI
(\ref{LMI.directed}), to obtain the matrix $Y$, then compute $\alpha_1$ and
$\alpha_2$;
(b) choose $\mu_i >0$ and $\omega >0$ to compute $\rho_i>0$;
(c) choose $\sigma_i \in (0, \rho_{\min})$;
(d) based on the desired upper bound $\Delta$, select $\gamma$, see
(\ref{bounded_e_i.directed});
(e) Lastly, choose $\nu_i$. Note that the term $\nu_i e^{-\sigma_i t}$
in (\ref{event condition TH2}) governs the triggering
threshold during the initial stage of the tracking process. Thus it
determines the frequency of communication events during this stage.
The value of  $\nu_i$ depends on the selected $\sigma_i$. If $\sigma_i$ is
large, then typically a relatively large $\nu_i$ must be chosen to ensure
the communication events occur less frequently.
\end{remark}

\section{Event-triggered leader-following control under an undirected
graph $\mathcal{G}$}
\label{main.1}

Although the 
problem for an undirected
$\mathcal{G}$ can be regarded as a
special case of the problem 
in Section~\ref{main.2}, an independent
derivation is of interest, which uses the symmetry of the matrix
$\mathcal{L}+G$. Accordingly, a different event
triggering condition is proposed for this case. 


\begin{theorem}\label{T1}
Let $R=R'>0$, $Q=Q'> 0$ be given matrices.
Suppose there exists a matrix $ Y= Y'> 0, Y \in \Re ^{n\times n}$, solving the following Riccati inequality
\begin{align}
\label{LMI.undirected}
YA + A'Y - 2 \underline \lambda YBR^{-1}B'Y +  Q \le 0,
\end{align}
where $\underline \lambda= \lambda_{\min} (\lambda_i)$ and $\lambda_i$ are the eigenvalues of $\mathcal{L}+G$.
Then under the control law (\ref{controller}) with $K=-R^{-1}B'Y$ the
system (\ref{agents dymamic}), (\ref{leader dymamic}) achieves the leader-follower tracking property of the form (\ref{defined tracking consensus})
with $\Delta =  N \gamma(\rho\lambda_{\min} (Y)\lambda_{\min} (F))^{-1}$,
if the communication events are triggered at
\begin{align}
\label{event condition TH1}
t_k^i=\inf \Big \{t>t_{k-1}^i \colon  &\|z_i\| \|s_i\|
\ge \frac{ \mu_i z_i' Q z_i + \nu_i e^{-\sigma_i t} + \gamma}{2\varpi_2} \Big\};
\end{align}
here $\varpi_2 = \lambda_{\max} (\mathcal {L} +G) \lambda_{\max} (YBR^{-1}B'Y)$, $\mu_i$, $\nu_i$, $\sigma_i$ and $\gamma$ are positive constants
chosen so that $0<\mu_i<1$, $\nu_i>0$, $\gamma>0$ and $\sigma_i\in (0,\rho)$,
where $\rho= (1 - \mu_{\max}) \lambda_{\min} (Q)/\lambda_{\max} (Y)$,
$\mu_{\max}= \max \limits_i \mu_i$.
In addition under this control law, Zeno behavior is ruled out:
\begin{align}
\inf_{k}(t_{k+1}^i - t_k^i)  \ge  \frac{1}{\|A\|} \ln \Big(1 + \frac{\|A\|\gamma}{ 2 \varpi_2 \bar\eta \hbar^2} \Big);
\label{inter-event-lower.bound}
\end{align}
here $\hbar= \sqrt{h/\lambda_{\min}(Y)}$, $h$ is defined in (\ref{VO.zeta1}) below.
\end{theorem}

\begin{remark}  \label{Remark2} \em
The Riccati inequality (\ref{LMI.undirected}) in this theorem is similar to the Riccati inequality employed in \cite{Garcia2014}. However,  our condition (\ref{LMI.undirected}) depends on the smallest eigenvalue $\underline \lambda$ of the matrix $\mathcal{L}+G$. In contrast, in \cite{Garcia2014} the second smallest eigenvalue of the graph Laplacian matrix is required to build the consensus algorithm.
When the graph topology is completely known at each node,  $\underline \lambda$
can be readily computed. But even when the graph $\mathcal{G}$ is not known at
each node, $\underline \lambda$ can be estimated in a decentralized
manner~\cite{Franceschelli2013}. Errors between the true eigenvalue
$\underline \lambda$ and its estimate $\underline {\hat\lambda}$ can be accommodated by replacing (\ref{LMI.undirected}) with a slightly more conservative condition. Suppose $|\underline \lambda -\underline{\hat\lambda}| < \varrho_1$, then the following Riccati inequality can be used in lieu of (\ref{LMI.undirected}):
\begin{align*} 
YA + A'Y - 2 (\underline {\hat\lambda} - \varrho_1)  YBR^{-1}B'Y +  Q \le 0.
\end{align*}
\end{remark}

\emph{Proof of Theorem~\ref{T1}:}
The proof is similar to the proof of Theorem~\ref{T2}
except for the procedure of obtaining an upper bound of
$z_i(t)$. Therefore, we only outline the proof of boundedness of
$z_i(t)$.
The closed loop system
consisting of error dynamics (\ref{error dymamic with norm}) is represented as
\begin{align}\label{large error dymamic}
\dot{\varepsilon}=(I_N\otimes A+ (\mathcal {L} +G)\otimes BK)\varepsilon + (I_N\otimes BK)s,
\end{align}
where as before $\varepsilon=[\varepsilon_1'~\ldots~\varepsilon_N']'$ and $s=[s_1'~\ldots~s_N']'$.

It follows from~\cite{Hong2006} that all the eigenvalues of matrix $\mathcal {L} +G$ ar positive.
Let $T\in \Re^{N\times N}$ be an orthogonal matrix such that $T^{-1}(\mathcal {L} +G)T=\Lambda =\mathrm{diag}\{\lambda_1, \ldots, \lambda_N\}$. Also, let $\zeta=(T^{-1}\otimes I_n)\varepsilon$,
$\zeta=[\zeta_1'~\ldots~\zeta_N']'$. Using this coordinate transformation, the system
(\ref{large error dymamic}) can be represented in terms of $\zeta$ and $s$, as
\begin{align}
\label{error dymamic transformation}
\begin{split}
 \dot{\zeta} =& \big({I_N\otimes A + \Lambda \otimes (BK)}\big)\zeta + (T^{-1}\otimes (BK))s.
\end{split}
\end{align}
Consider the following Lyapunov
function candidate for the system (\ref{error dymamic transformation}),
$V(\zeta)=\zeta' (\Lambda^2 \otimes Y)\zeta$.
Using (\ref{LMI.undirected}), the
coordinate transformation $\zeta=(T^{-1}\otimes I_n)\varepsilon$, the
identity $z=\big((\mathcal{L} +G)\otimes I_n\big)\varepsilon,
z=[z_1', \ldots, z_N']'$, and  condition (\ref{event
  condition TH1}) we can show that
on every interval $[t_k^i,~t_{k+1}^i)$,
\begin{align}
V(\zeta)  \le&  e^{- \rho t} \Big(V(\zeta(0))- \sum_{i=1}^{N} \frac{\nu_i}{\rho-\sigma_i} - \frac{N\gamma}{\rho}\Big)  \nonumber \\
 &+ \sum_{i=1}^{N} e^{- \sigma_i t}\frac{\nu_i}{\rho-\sigma_i} + \frac{N
   \gamma}{\rho}  \nonumber \\
\label{VO.zeta1}
\le & V(\zeta(0))+ \sum_{i=1}^{N}\frac{\nu_i}{\rho-\sigma_i} + \frac{N \gamma}{\rho} =h.
\end{align}
$V(\zeta)$ can be expressed in terms of $\varepsilon$ using the
inverse transformation $\zeta= (T^{-1} \otimes I_n) \varepsilon$ and $z= ((\mathcal{L}+G) \otimes I_n) \varepsilon$
\begin{align}
\label{lyap-consensus}
&V(\zeta)= \varepsilon' ((\mathcal{L}+G)^2 \otimes Y) \varepsilon = z' (I_N \otimes Y) z. 
\end{align}
It then follows from (\ref{lyap-consensus}) that
$\lambda_{\min}(Y)\sum_{i=1}^{N}  \| z_i(t)\|^2 
\le V(\zeta).$
The rest of the proof of this theorem is similar to the proof of Theorem~\ref{T2} and is omitted for brevity.
\hfill$\Box$

\begin{remark} \em
It can be shown that the observation made in Remark~\ref{remark
  Th.directed} applies in this case as well.
The parameters in the event triggering conditions (\ref{event
  condition TH1}) can be selected following a process similar to that
outlined in Remark~\ref{remark event.paramters.directed}.
\end{remark}

\section{Generation of the combinational state}
\label{algorithm}
To implement the event triggering conditions (\ref{event condition TH2})
and (\ref{event condition TH1}) in Theorems~\ref{T2} and ~\ref{T1}, the combinational state $z_i(t)$ must be known at all times.
We now describe how node $i$ can generate $z_i(t)$ continuously
using only discrete communications from its neighbors at event times. This
eliminates the need for agent $i$ to monitor and communicate with its neighbors
continuously.

According to
(\ref{agents dymamic}), for $t \in [t_k^i, t_{k+1}^i)$, the
dynamics of $x_i(t)$ and $x_j(t)$, $j \in N_i$, on this interval can be
expressed as
\begin{eqnarray}
\label{x_i}
\lefteqn{x_i(t)= e^{A(t-t_k^i)} x_i(t_k^i) - \int_{t_k^i}^{t} e^{A(t-\tau)} BK
z_i(t_k^i) d\tau,} && \\
\label{x_j}
\lefteqn{x_j(t)= e^{A(t-t_k^i)} x_j(t_k^i) - \int_{t_k^i}^{\hat t_j}
  e^{A(t-\tau)} BK z_j (t_l^j) d\tau} && \nonumber \\
&&\hspace{2ex}- \sum \limits_{m \colon t_k^i < t_ m^j <t} \int_{t_m^j}^{\min (t, t_{m+1}^j)} e^{A(t-\tau)} BK z_j (t_m^j) d\tau,\\
\lefteqn{\hat t_j = \left\{
  \begin{array}{ll}
    t_{l+1}^j, & \textrm{if $j$ has at least one event on $[t_k^i, t)$}, \\
    t, & \textrm{otherwise},
  \end{array}
\right.} && \nonumber
\end{eqnarray}
where $t_{l+1}^j= \min (t_m^j: t_m^j \in [t_k^i, t))$.
Equation (\ref{x_j}) accounts for the fact that agent $j$ may experience
several events at times $t_m^j$, $m=l+1$, $\ldots$, within the time
interval $[t_k^i, t_{k+1}^i)$. When $[t_k^i, t)$ contains no event
triggered by agent $j$, the last term in (\ref{x_j}) vanishes. Similarly, the dynamics of the tracking error $\varepsilon_i(t)$ can be
expressed as
\begin{align}
\label{varepsilon_i}
 \varepsilon_i(t)=  e^{A(t-t_k^i)} \varepsilon_i(t_k^i) + \int_{t_k^i}^{t} e^{A(t-\tau)} BK z_i(t_k^i) d\tau.
\end{align}
Using the notation $\Phi(t,t')= \int_{t}^{t'} e^{A(t'-\tau)}  BK  d \tau$, it
follows from (\ref{define z_i}), (\ref{x_i}), (\ref{x_j}) and
(\ref{varepsilon_i}) that
\begin{align}
\label{z_i 1}
 &z_i(t)= e^{A(t-t_k^i)} z_i(t_k^i)  + g_i \Phi(t_k^i, t) z_i(t_k^i)  \nonumber \\
 &+ \sum\limits_{j\in N_i}  \sum \limits_{m \colon t_k^i < t_ m^j <t} \Phi(t_m^j, \min (t, t_{m+1}^j)) (z_i(t_k^i) - z_j(t_m^j)) \nonumber \\
 &+ \sum\limits_{j\in N_i} \Phi(t_k^i, \hat t_j) (z_i(t_k^i) - z_j(t_l^j)).
\end{align}
According to (\ref{z_i 1}), to generate $z_i(t)$,
agent $i$ must know $z_i (t_k^i) $ and $z_j(t_m^j)$, $t_k^i < t_m^j <t$. It
has $z_i(t_k^i)$ in hand and thus it must
only receive $z_j(t_m^j)$ when an event is
triggered at node $j$ during $[t_k^i, t_{k+1}^i)$. To ensure this, we
propose an algorithm to allow every agent
to compute its combinational state and broadcast it to its neighbors at
time instants determined by its triggering condition. This algorithm has
a noteworthy feature that follows
from (\ref{z_i 1}) in that only one-way communications occur between
the neighboring agents at the triggering time, even
when the graph $\mathcal{G}$ is undirected. 

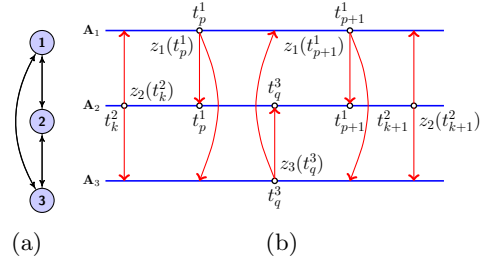
\begin{figure}
\centering
\subfloat[]{
 \label{Undirected.CG.1}
    \scalebox{0.5}{
  \begin{tikzpicture}[->,>=stealth',shorten >=1pt,auto,node distance=2.1cm,
   thick,main node/.style={circle,fill=blue!20,draw,font=\sffamily\bfseries},]
    \node[main node] (1) {1};
   \node[main node] (2) [below of=1] {2};
   \node[main node] (3) [below of=2] {3};
   \path[every node/.style={font=\sffamily\small}]
        (1) edge node {} (2)   (2) edge node {} (1)
        (2) edge node {} (3)   (3) edge node {} (2)
        (3) edge  [bend left] node {} (1)    (1) edge  [bend right] node {} (3) ;
 \end{tikzpicture}
 }
 }
\subfloat [] {\label{Undirected.CG.2}
\scalebox{0.5}{
\begin{tikzpicture}
[decoration={markings,%
 mark=at position .999 with {\arrow[red,line width=2pt]{>}},%
 }]
    \filldraw[fill=white,line width=1pt](-4.5,2)node[left]{$\textrm{\textbf{A}$_1$}$};
    \filldraw[fill=white,line width=1pt](-4.5,0)node[left]{$\textrm{\textbf{A}$_2$}$};
   \filldraw[fill=white,line width=1pt](-4.5,-2)node[left]{$\textrm{\textbf{A}$_3$}$};

    \draw[blue,line width=1.2pt](-4.5,2)--(4.5,2);
    \filldraw[fill=white,line width=1pt](-2,2)circle(.07cm)node[above]{\large $t_{p}^1$}node[below left]{\large $z_1(t_{p}^1)$};
    \filldraw[fill=white,line width=1pt](2,2)circle(.07cm)node[above]{\large $t_{p+1}^1$}node[below left]{\large $z_1(t_{p+1}^1)$};
    \draw[blue,line width=1.2pt](-4.5,0)--(4.5,0);
    \filldraw[fill=white,line width=1pt](-4,0)circle(.07cm)node[below left]{\large $t_{k}^2$}node[above right]{\large $z_2(t_{k}^2)$};
    \filldraw[fill=white,line width=1pt](3.7,0)circle(.07cm)node[below left]{\large $t_{k+1}^2$}node[below right]{\large $z_2(t_{k+1}^2)$};

     \filldraw[fill=white,line width=1pt](-2,0)circle(.07cm)node[below]{\large $t_{p}^{1}$};
    \filldraw[fill=white,line width=1pt](2,0)circle(.07cm)node[below]{\large $t_{p+1}^{1}$};
    \filldraw[fill=white,line width=1pt](0,0)circle(.07cm)node[above]{\large $t_{q}^{3}$};

      \draw[blue,line width=1.2pt](-4.5,-2)--(4.5,-2);
     \filldraw[fill=white,line width=1pt](0,-2)circle(.07cm)node[below]{\large $t_{q}^3$}node[above right]{\large $z_3(t_{q}^3)$};
  \draw[
     red, thick,postaction={decorate} ]%
     plot[smooth] coordinates {(-2,1.95) (-1.65,1) (-1.5,0) (-1.65,-1) (-2,-2)};

  \draw[
     red, thick,postaction={decorate} ]%
     plot[smooth] coordinates {(2,1.95) (2.35,1) (2.485,0) (2.35,-1) (2,-2)};

\coordinate (A12) at (-2,1.95);
\coordinate (B12) at (-2,0.05);
\draw [red, thick, postaction={decorate}] (A12) to  (B12);

\coordinate (A12) at (2,1.95);
\coordinate (B12) at (2,0.05);
\draw [red, thick, postaction={decorate}] (A12) to  (B12);

\coordinate (A211) at (-4,2);
\coordinate (B211) at (-4,0.05);
\draw [red, thick, postaction={decorate}] (B211) to  (A211);

\coordinate (C231) at (-4,-2);
\coordinate (B231) at (-4,-0.05);
\draw [red, thick, postaction={decorate}] (B231) to  (C231);

\coordinate (A212) at (3.7,2);
\coordinate (B212) at (3.7,0.05);
\draw [red, thick, postaction={decorate}] (B212) to  (A212);

\coordinate (C232) at (3.7,-2);
\coordinate (B232) at (3.7,-0.05);
\draw [red, thick, postaction={decorate}] (B232) to  (C232);

  \draw[
     red, thick,postaction={decorate} ]%
     plot[smooth] coordinates {(0,-1.95) (-0.35,-1) (-0.5,0) (-0.35,1) (0,2)};

\coordinate (C32) at (0,-1.95);
\coordinate (B32) at (0,-0.05);
\draw [red, thick, postaction={decorate}] (C32) to  (B32);
\end{tikzpicture}
}
 }
 \caption{Communication between followers in an undirected
   network: (a) The graph; (b) Communication events.}
  \label{Undirected.CG}
\end{figure}

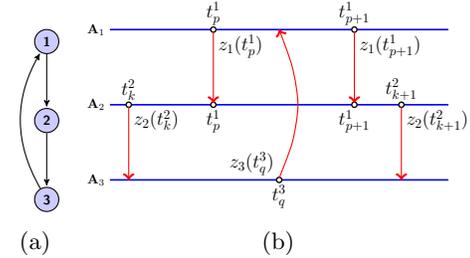
\begin{figure}
\centering
\subfloat[]{
 \label{Directed.CG.1}
    \scalebox{0.5}{
  \begin{tikzpicture}[->,>=stealth',shorten >=1pt,auto,node distance=2.1cm,
   thick,main node/.style={circle,fill=blue!20,draw,font=\sffamily\bfseries},]
    \node[main node] (1) {1};
   \node[main node] (2) [below of=1] {2};
   \node[main node] (3) [below of=2] {3};
   \path[every node/.style={font=\sffamily\small}]
        (1) edge node {} (2)
        (2) edge node {} (3)
        (3) edge  [bend left] node {} (1);
 \end{tikzpicture}
 }
 }
\subfloat [] {\label{Directed.CG.2}
\scalebox{0.5}{
\begin{tikzpicture}
[decoration={markings,%
 mark=at position .999 with {\arrow[red,line width=2pt]{>}},%
 }]
    \filldraw[fill=white,line width=1pt](-4.5,2)node[left]{$\textrm{\textbf{A}$_1$}$};
    \filldraw[fill=white,line width=1pt](-4.5,0)node[left]{$\textrm{\textbf{A}$_2$}$};
   \filldraw[fill=white,line width=1pt](-4.5,-2)node[left]{$\textrm{\textbf{A}$_3$}$};

    \draw[blue,line width=1.2pt](-4.5,2)--(4.5,2);
    \filldraw[fill=white,line width=1pt](-1.75,2)circle(.07cm)node[above]{\large $t_{p}^1$}node[below right]{\large $z_1(t_{p}^1)$};
      \filldraw[fill=white,line width=1pt](2,2)circle(.07cm)node[above]{\large $t_{p+1}^1$}node[below right]{\large $z_1(t_{p+1}^1)$};
    \draw[blue,line width=1.2pt](-4.5,0)--(4.5,0);
    \filldraw[fill=white,line width=1pt](-4,0)circle(.07cm)node[above]{\large $t_{k}^2$}node[below right]{\large $z_2(t_{k}^2)$};
    \filldraw[fill=white,line width=1pt](3.25,0)circle(.07cm)node[above]{\large $t_{k+1}^2$}node[below right]{\large $z_2(t_{k+1}^2)$};

     \filldraw[fill=white,line width=1pt](-1.75,0)circle(.07cm)node[below]{\large $t_{p}^{1}$};
    \filldraw[fill=white,line width=1pt](2,0)circle(.07cm)node[below]{\large $t_{p+1}^{1}$};

     \draw[blue,line width=1.2pt](-4.5,-2)--(4.5,-2);
     \filldraw[fill=white,line width=1pt](0,-2)circle(.07cm)node[below]{\large $t_{q}^3$}node[above left]{\large $z_3(t_{q}^3)$};

\coordinate (A12) at (-1.75,1.95);
\coordinate (B12) at (-1.75,0.05);
\draw [red, thick, postaction={decorate}] (A12) to  (B12);

\coordinate (A12) at (2,1.95);
\coordinate (B12) at (2,0.05);
\draw [red, thick, postaction={decorate}] (A12) to  (B12);

\coordinate (C231) at (-4,-2);
\coordinate (B231) at (-4,-0.05);
\draw [red, thick, postaction={decorate}] (B231) to  (C231);

\coordinate (C232) at (3.25,-2);
\coordinate (B232) at (3.25,-0.05);
\draw [red, thick, postaction={decorate}] (B232) to  (C232);

  \draw[
     red, thick,postaction={decorate} ]%
     plot[smooth] coordinates {(0,-1.95) (0.35,-1) (0.5,0) (0.35,1) (0,2)};
\end{tikzpicture}
}
 }
 \caption{Communication between followers in a directed
   network: (a) The graph; (b) Communication events.}
  \label{Directed.CG}
\end{figure}

Before presenting the algorithm formally, let us illustrate using (\ref{z_i
  1}) with an example involving three
agents, $\textbf{A}_1$, $\textbf{A}_2$ and $\textbf{A}_3$; see
Fig.~\ref{Undirected.CG} and~\ref{Directed.CG}. E.g., consider the timeline in
Fig.~\ref{Undirected.CG.2}. According to the timeline of $\textbf{A}_2$, an
event has been triggered for $\textbf{A}_2$ at time $t_k^2$. Until it
receives
communications from the neighbors, $\textbf{A}_2$ computes $z_2(t)$ using
(\ref{z_i 1}) with the information $z_1(t_{p-1}^1)$ and
$z_3(t_{q-1}^3)$ received from $\textbf{A}_1$ and
$\textbf{A}_3$ prior time $t_k^2$. This information is used in the third line of (\ref{z_i
  1}). Note that $\hat
t_1=\hat t_3=t$ until $\textbf{A}_2$ receives the first message;
the terms in the second line are  zero until then.
At time $t_p^1$, an event occurs at node 1, and $\textbf{A}_2$
receives the
value of $\textbf{A}_1$'s combinational state, $z_1(t_p^1)$. From
this time on, it starts using $\Phi(t_p^{1},t)(z_2(t_k^2)-z_1(t_p^1))$
in (\ref{z_i 1}), until the next message arrives, this time from
$\textbf{A}_3$.
Overall during the interval $[t_k^2,t_{k+1}^2)$, $\textbf{A}_2$ receives
$z_1(t_p^1)$ and $z_1(t_{p+1}^1)$ from $\textbf{A}_1$ and $z_3(t_q^3)$ from $\textbf{A}_3$, which it uses 
in (\ref{z_i 1}) 
to compute $z_2(t)$. When $\mathcal{G}$ is directed, $\textbf{A}_2$
computes $z_2(t)$ in the same manner, but it only receives $z_1(t_p^1)$ and
$z_1(t_{p+1}^1)$ from $\textbf{A}_1$, as shown in
Fig.~\ref{Directed.CG.2}.


We conclude this section by summarizing the algorithm for
generating the combinational state $z_i(t)$ at each node.

\begin{description}
 \item [\textbf{Initialization. }]
  \begin{enumerate}[(a)]
  \item Synchronize local clocks to set $t=0$ at each node, also set the event 
counter $k=0$, the local event time record $t_k^i=0$, and the local
measurement error $s_i(t)=0$.
  \item Receive $x_j(0)$ from all neighbors $j \in N_i$;
  \item Send $x_i(0)$ to agents $r$ such that $i \in N_r$;
  \item Compute $z_i(0)$ using the received $x_j(0)$, $j\in N_i$;
  \item Receive $z_j(0)$, $j \in N_i$, and send $z_i(0)$ to
    agents $r$ such that $i \in N_r$.
  \end{enumerate}
\item [\textbf{Do While}]
 (\ref{event condition TH2}) (if $\mathcal{G}$
   is directed) or (\ref{event condition TH1}) (if $\mathcal{G}$
   is undirected) is not satisfied:
\begin{enumerate}[(a)]
 \item
Compute $z_i(t)$ with the latest received $\hat z_j(t)$, $j \in N_i$, using
(\ref{z_i 1}), then update $s_i(t)$ using (\ref{measure error});
\end{enumerate}
\item [\textbf{Else}]\mbox{}
\begin{enumerate}[(a)]
 \item Advance the event counter $k=k+1$, and set $t_k^i=t$, $s_i(t)=0$;
 \item Set $z_i(t_k^i)=z_i(t)$ and send $z_i(t_k^i)$ to agents $r$ such that $i
   \in N_r$;
 \item Update the control signal $u_i=-K z_i(t_k^i) $.
 \end{enumerate}
\item [End]
\end{description}


As one can see, the algorithm uses only one-directional communications between
agents at event times: the information is received from $j\in N_i$ when an
event occurs at node $j$ and is sent to $r$, $i\in N_r$ when an
event occurs at node $i$, e.g., see Figs.~\ref{Undirected.CG.2} and~\ref{Directed.CG.2}.


\section{Example}\label{example}

\begin{figure}[t]
\centering
 \scalebox{0.8}{
\begin{tikzpicture}[
    media/.style={font={\footnotesize\sffamily}},
    wave/.style={
        decorate,decoration={snake,post length=1.4mm,amplitude=2mm,
        segment length=2mm},thick},
    interface/.style={
        postaction={draw,decorate,decoration={border,angle=-45,
                    amplitude=0.3cm,segment length=2mm}}},
    gluon/.style={decorate, draw=black,
        decoration={coil,amplitude=4pt, segment length=5pt}},scale=0.7
    ]
    \draw[blue,line width=1pt](-5.2,0)--(0,0);
     \draw[blue,line width=1pt](1.1,0)--(2.3,0);
    \draw[dotted, blue,thick](0.15,0)--(0.9,0);

    \draw[dashed,black](-5,-2.3)--(-5,0);
    \draw[dashed,black](-2,-2.3)--(-2,0);
    \draw[dashed,black](2,-2.3)--(2,0);

    \draw[black](0:-5cm)--(34:-5.1cm)node[midway]{\scriptsize $~~l$};
    \path (-5,0)++(-84:2.2cm)node{\scriptsize  $~\alpha_0$};
    \draw[->] (-5,-1.5) arc (-90:-63:.75cm);

    \draw[black](0:-2cm)--(71:-3.4cm);
    \path (-2,0)++(-84:2.2cm)node{\scriptsize  $~\alpha_1$};
    \draw[->] (-2,-1.5) arc (-90:-63:.75cm);

    \draw[black](2,0)--(3.0,-3);
    \path (2,0)++(-82:2.2cm)node{\scriptsize  $~\alpha_{20}$};
    \draw[->] (2,-1.5) arc (-90:-63:.9cm);

\draw[dotted, black,thick](0.15,-2)--(0.9,-2);

    \filldraw[fill=white,line width=1pt](-4.2,-3)circle(.2cm)node[left]{\scriptsize $Leader~$};
       \filldraw[fill=white,line width=1pt](-1.2,-3)circle(.2cm)node[left]{\scriptsize $u_1 \rightarrow~$};
          \filldraw[fill=white,line width=1pt](3.0,-3)circle(.2cm)node[left]{\scriptsize $u_{20} \rightarrow~$};
\end{tikzpicture}
}
\caption{The system consisting of twenty pendulums and the leader pendulum.}
\label{pendulums}
\end{figure}
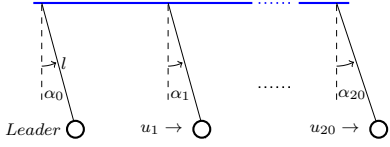

Consider a system consisting of twenty
identical pendulums. Each pendulum is subject to an
input as shown in Fig.~\ref{pendulums}. The dynamic of the $i$-th pendulum
is governed by the following linearized equation
\begin{align}
\label{dynamic of pedulums}
ml^2\ddot{\alpha}_i=&-mgl\alpha_i-u_i, \quad i=1,\ldots, 20,
\end{align}
where $l$ is the length of the pendulum, $g=9.8$~m/s$^2$ is the gravitational
acceleration constant, $m$ is the mass of each pendulum and $u_i$ is the
control torque (realized using a DC motor). In
addition, consider the leader pendulum which is
identical to those given and whose dynamic is described
by the linearized equation
\begin{equation}
\label{leader pedulums}
ml^2\ddot{\alpha}_0=-mgl\alpha_0.
\end{equation}
Choosing the state vectors as $x_i=(\alpha_i, \dot{\alpha}_i)$,
$i=0,\ldots, 20$, equations (\ref{dynamic of pedulums}) and (\ref{leader
  pedulums}) can be written in the form of (\ref{agents dymamic}),
(\ref{leader dymamic}), where
$A={\scriptsize\left[\begin{array}{cc}
                          0  &  1  \\
                        -g/l &  0
                        \end{array}\right]}$, $B={\scriptsize\left[\begin{array}{c}
                          0    \\
                        -1/(ml^2)
                        \end{array}\right]}$.
In this example, we let $m=1$~kg, $l=1$~m. 


%

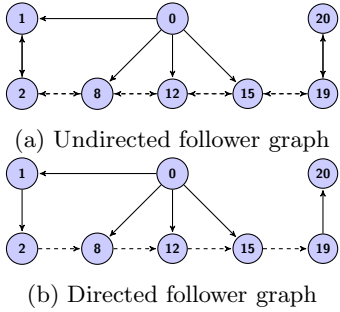
\begin{figure}[t]
\centering
\subfloat[][Undirected follower graph]{
 \label{CG.1}
  \scalebox{0.5}{
 \begin{tikzpicture}[->,>=stealth',shorten >=1pt,auto,node distance=2cm,
   thick,main node/.style={circle,fill=blue!20,draw,font=\sffamily\bfseries},]

   \node[main node] (2) {~1~~};
   \node[main node] (3) [below of=2]{~2~~};
   \node[main node] (4) [right of=3] {~8~~};
   \node[main node] (5) [right of=4] {12};
   \node[main node] (6) [right of=5] {15};
   \node[main node] (7) [right of=6] {19};
   \node[main node] (8) [above of=7] {20};
   \node[main node] (1) [above of=5]{~0~~};

   \path[every node/.style={font=\sffamily\small}]
   (1) edge node {}  (2)
   (1) edge node {}  (4)
   (1) edge node {}  (5)
   (1) edge node {}  (6)
   (2) edge node {}  (3)
   (3) edge node {}  (2)
   (7) edge node {}  (8)
   (8) edge node {}  (7);

 \path[draw,dashed] (3)--(4);
 \path[draw,dashed] (4)--(3);

 \path[draw,dashed] (4)--(5);
 \path[draw,dashed] (5)--(4);

 \path[draw,dashed] (5)--(6);
 \path[draw,dashed] (6)--(5);

  \path[draw,dashed] (6)--(7);
 \path[draw,dashed] (7)--(6);

 \end{tikzpicture}
 }
 } \quad
\subfloat [][Directed follower graph] {\label{CG.2}
 \scalebox{0.5}{
  \begin{tikzpicture}[->,>=stealth',shorten >=1pt,auto,node distance=2cm,
   thick,main node/.style={circle,fill=blue!20,draw,font=\sffamily\bfseries}]


   \node[main node] (2) {~1~~};
   \node[main node] (3) [below of=2]{~2~~};
   \node[main node] (4) [right of=3] {~8~~};
   \node[main node] (5) [right of=4] {12};
   \node[main node] (6) [right of=5] {15};
   \node[main node] (7) [right of=6] {19};
   \node[main node] (8) [above of=7] {20};
   \node[main node] (1) [above of=5]{~0~~};

   \path[every node/.style={font=\sffamily\small}]
   (1) edge node {}  (2)
   (1) edge node {}  (4)
   (1) edge node {}  (5)
   (1) edge node {}  (6)
   (2) edge node {}  (3)
   (7) edge node {}  (8);
 \path[draw,dashed] (3)  -- (4);
 \path[draw,dashed] (4)  -- (5);
 \path[draw,dashed] (5)  -- (6);
 \path[draw,dashed] (6)  -- (7);

 \end{tikzpicture}
 }
 }
 \caption{Communication graphs for the example.}
  \label{CG}
\end{figure}

Both undirected and directed follower graphs $\mathcal{G}$ are considered in the example, shown in
Fig.~\subref*{CG.1} and~\subref*{CG.2}, respectively. According to
Fig.~\ref{CG}, in both cases agents $1$, $8$, $12$ and $15$ measure
the leader's state, however in the graph in Fig.~\subref*{CG.2}
follower $i$ is restricted to receiving information from follower $i-1$ only,
whereas in Fig.~\subref*{CG.1}, it can receive information from both $i-1$
and $i+1$.

We implemented four simulations to compare the results proposed in this
paper and also to compare them with the results in~\cite{Zhu2015}. The
directed graph in Fig.~\subref*{CG.2} was 
employed to illustrate Theorem~\ref{T2} in Simulation 1.
In Simulation 2, we implemented the controller designed using Theorem~\ref{T1} with the undirected graph in Fig.~\subref*{CG.1}.
We applied Theorem~\ref{T2} using the same undirected graph in Fig.~\subref*{CG.1} in Simulation~3.
In Simulation 4, we applied the event-based control strategy proposed
in Theorem 3~\cite{Zhu2015}, also using the directed graph in
Fig.~\subref*{CG.2}. Out of the results in~\cite{Zhu2015},
we chose Theorem 3 for comparison, because it has a
way to avoid continuous communications between the followers; this allows
for a fair comparison with our methods.

In the first three simulations, we aimed to restrict the predicted upper bound on
the tracking error to $\Delta \le 0.05$. In the design, we chose the same
$Q$ matrix and adjusted $R$ to obtain the same control gains
in the three simulations. The parameters of the triggering conditions
(\ref{event condition TH2}) and (\ref{event condition TH1}) and the design
parameters were set as shown in Table~\ref{event.simulation.results}. 
In Simulation~4, using the same matrix $Q$, we computed the control
gain $K=[2.63,~7.24]$ and also chose the 
parameters required by Theorem~3 of \cite{Zhu2015} as follows: $h=5$,
$\beta_1=0.1$, $\beta_2=0.15$, $\gamma=0.2$ and $\tau=0$ (see
\cite{Zhu2015} for the definition of these parameters). In all 
simulations we endeavoured to 
achieve the least number of communications events. 

\begin{figure}
\centering
\subfloat[][Simulation 1]{\label{tracking.errors.1}
\includegraphics[width=0.8\columnwidth]{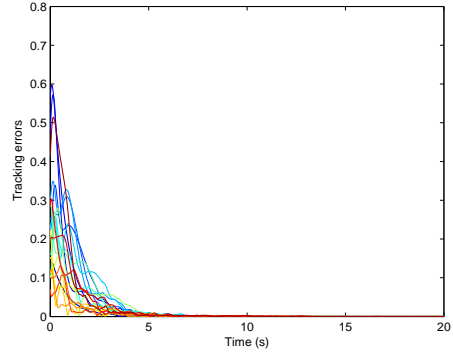}
 } \quad
\subfloat [][Simulation 2] {\label{tracking.errors.2}
\includegraphics[width=0.8\columnwidth]{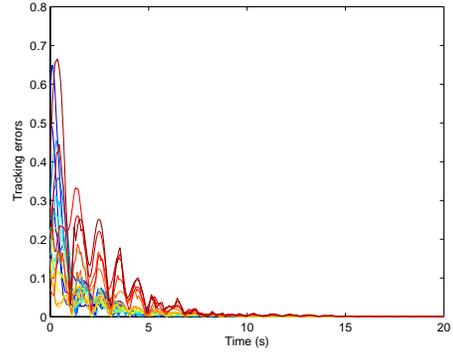}
 }\quad
 \subfloat [][Simulation 3] {\label{tracking.errors.3}
\includegraphics[width=0.8\columnwidth]{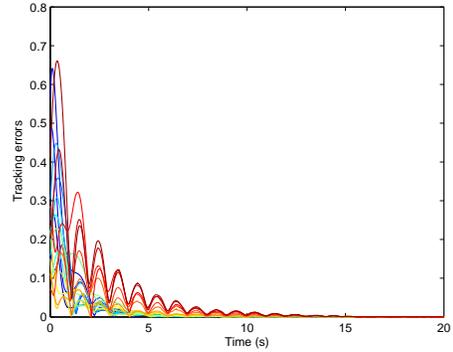}
 }\quad
 \subfloat[][Simulation 4]{\label{tracking.errors.4}
\includegraphics[width=0.8\columnwidth]{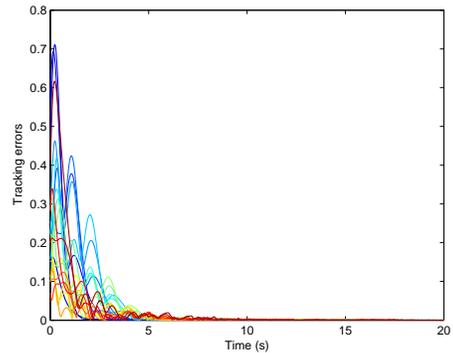}
 }
  \caption{Tracking errors $\|\varepsilon_i\|$.}
  \label{tracking.errors}
\end{figure}

The simulation results achieved in Simulations 1-3 are shown in
Table~\ref{event.simulation.results}. 
In the table, $J_{[18,20]}$ denotes the maximum actual
tracking error $ \sum_{i=1}^{N}  \|\varepsilon_i(t) \|^2$ observed over the time interval $t \in [18, 20]$,
$t_{[0,20]}$, $t_{[0,10]}$, $t_{[10,20]}$ and $E_{[0,20]}$, $E_{[0,10]}$,
$E_{[10,20]}$ represent the minimum inter-event intervals and the total number
of events occurred in the system on the time intervals $[0,20]$, $[0,10]$, $[10,20]$, respectively. 
The corresponding results of Simulation~4 are $J_{[18,20]}=1.8778\times 10^{-6}$, $t_{[0,20]}=21.1$~ms, $t_{[0,10]}=21.1$~ms, $t_{[10,20]}=101.5$~ms, $E_{[0,20]}=2857$, $E_{[0,10]}=1767$ and
$E_{[10,20]}=1090$. 
The tracking errors are shown in Fig.~\ref{tracking.errors}, which
illustrates that all four event-triggered tracking control laws enable all
the followers to synchronize to the leader. 

\begin{table}[!tb]
\centering
\addtolength{\tabcolsep}{-3.5pt}
\caption{The design parameters and simulation results.}  \label{event.simulation.results}
\begin{tabular}{ccccccc}

\toprule
\multicolumn{1}{c}{}     & \begin{tabular}[c]{@{}c@{}}Simulation 1 \\ (Theorem~\ref{T2} \\and Fig.~\ref{CG.2})\end{tabular} & \begin{tabular}[c]{@{}c@{}}Simulation 2 \\ (Theorem~\ref{T1}\\ and Fig.~\ref{CG.1})\end{tabular} & \begin{tabular}[c]{@{}c@{}}Simulation 3 \\ (Theorem~\ref{T2}\\and Fig.~\ref{CG.1})\end{tabular} \\
 \midrule
 $Q$           &  $\begin{bmatrix}10.59     & 0.42  \\0.42    & 1.05   \end{bmatrix}$
             &  $\begin{bmatrix}10.59     & 0.42  \\0.42    & 1.05   \end{bmatrix}$
             &  $\begin{bmatrix}10.59     & 0.42  \\0.42    & 1.05   \end{bmatrix}$ \\
 $R$           & $1.1394$    &  $0.1$   &  $1.5405$            \\
 $K$         &$[5.23~ 13.08]$    &  $[5.23~13.08]$    &  $[5.23~13.08]$   \\
$\alpha$     &$0.0877$           &  $-$               &  $0.0649$          \\
$\omega$     & 0.001             &  $-$               &  $0.001$            \\
$\mu_i$      & 0.1               &  $0.1$             &  $0.1$             \\
$\sigma_i$   & $0.5025$          & $0.7198$ & $0.3007$          \\
$\nu_i$      & $2.5$             & $2$ &  $1.2$              \\
$\gamma$     & $ 2.9769\times10^{-5}$  & $7.9990\times10^{-6}$ &  $3.1507\times10^{-6}$ \\
$\Delta$     & $ 0.0462$          & $0.0462$           &  $0.0462$           \\
$J_{[18,20]}$&$3.4524\times10^{-7}$&$3.5259\times10^{-6}$&  $1.5518\times10^{-6}$ \\
$t_{[0,20]}$ & $6.2$~ms          &  $5.9$~ms          &  $1.9$~ms            \\
$t_{[0,10]}$ & $6.2$~ms          &  $5.9$~ms          &  $1.9$~ms            \\
$t_{[10,20]}$ & $30.7$~ms        &  $24.9$~ms         &  $26.0$~ms           \\
$E_{[0,20]}$  & $2240$           &  $2091$            &  $2979$              \\
$E_{[0,10]}$  & $1285$           &  $1217$            &  $1993$              \\
$E_{[10,20]}$ & $955$           &  $874$              &  $986$              \\
\bottomrule
\end{tabular}
\end{table}

The first comparison was made between the techniques developed in this
paper for systems connected over directed 
and undirected graphs; these techniques were applied in Simulations 1 and
2, respectively. 
Although the minimum inter-event intervals in Simulation 2 were observed to
be smaller than those in Simulation 1, on average the events were triggered
less frequently in Simulation 2. This demonstrates that connecting the
followers into an undirected network and using the design scheme based on
Theorem~\ref{T1} may lead to 
some advantages in terms of usage of communication resources.

Next, we compared Simulations~2 and 3 using the same undirected follower
graph in Fig.~\subref*{CG.1} based on Theorems~\ref{T2} and \ref{T1}
developed in the paper. Compared with Simulation~2, more communication
events and smaller minimum inter-event intervals were observed in
Simulation~3. One possible explanation to this is because the method based
on Theorem~\ref{T1} takes an advantage of the symmetry property of the
matrix $\mathcal{L}+G$ of the undirected follower graph in the derivation. 

Finally, we compared Simulations 1 and 4 where we used the same directed
follower graph in Fig.~\subref*{CG.2} for both designs. Although compared
with the method of Theorem~3 of \cite{Zhu2015}, 
our method produced smaller minimum time intervals between the events, the
total number of events occurred during the simulation 
using our method was also smaller. We remind that we endeavoured to select
the simulation parameters and the controller gains for this simulation to
reduce the total number of events. We also tried to compare the
performance of the two methods by tuning the controller of~\cite{Zhu2015}
to almost the same gain as in Simulation 1, but the
results in Simulation~4 were even worse, producing a much greater number of
communication events ($E_{[0,20]}=5553$, $E_{[0,10]}=3100$ and
$E_{[10,20]}=2453$).

\section{Conclusions}
\label{conclusion}
The paper has studied the event-triggered leader-follower tracking control
problem for multi-agent systems. We have presented sufficient
conditions to guarantee that the proposed event-triggered control scheme
leads to bounded tracking errors. Furthermore, our results show that
by adjusting the parameters of the triggering condition, the upper bound on
the tracking errors guaranteed by these conditions can be tuned to a desired small value, at the expense of more frequent communications during an
initial stage of the tracking process. Such conditions have been derived
for both undirected and directed follower graphs. Also, we showed that the
proposed event triggering conditions do not lead to Zeno behavior
even if a tight accuracy requirement on the tracking errors is imposed.
In fact, with the proposed triggering rules, such tight accuracy
requirements do not impact the inter-event intervals after a sufficiently
large time. We also presented a computational algorithm which allows the
nodes to continuously generate the combinational state at every node which
is needed to implement these event triggering schemes. Thus, continuous
monitoring the neighboring states is avoided. The efficacy of the proposed
algorithm has been demonstrated using a simulation example. Future work will include the study of robustness of the proposed control scheme.

\section{Acknowledgements}
The authors thank the Associate Editor and
the Reviewers for their helpful and
constructive comments.


\begin{thebibliography}{1}

%
%
%

\bibitem{Ren2008}
W. Ren and R. W. Beard, {\em Distributed consensus in multi-vehicle cooperative
control}. London: Springer-Verlag, 2008.


\bibitem{Xie2009}
G. Xie, H. Liu, L. Wang, and Y. Jia, ``Consensus in networked multi-agent systems via sampled control: fixed topology case," Proc. ACC, 2009, pp.3902--3907.
\bibitem{Mazo2010}
M. Mazo Jr., A. Anta, and P. Tabuada, ``An ISS self-triggered implementation of linear controller," {\em Automatica}, \textbf{46}, 1310--1314, 2010.

\bibitem{Heemels2012}
W. P. M. H. Heemels, K. H. Johansson, and P. Tabuada, ``An introduction to
event-triggered and self-triggered control," Proc.  51st IEEE CDC, 2012, pp.3270--3285.

\bibitem{Tabuada2007}
P. Tabuada, ``Event-triggered real-time scheduling of stabilizing control tasks," {\em IEEE Trans. Autom. Contr.},  \textbf{52}, 1680--1685, 2007.

\bibitem{Lunze2010}
J. Lunze and D. Lehmann, ``A state-feedback approach to event-based control,"
{\em Automatica}, \textbf{46}, 211--215, 2010.


\bibitem{Dimarogonas2009}
D. V. Dimarogonas and K. H. Johansson,``Event-triggered control
for multi-agent systems'', Proc. 48th IEEE CDC - 28th CCC, 2009, pp.7131--7136.

\bibitem {Wang2011}
X. Wang and M. Lemmon, ``Event-triggering in distributed networked control systems," {\em IEEE Trans. Autom. Contr.}, \textbf{56}, 586--601, 2011.

\bibitem{Dimarogonas2012}
D. V. Dimarogonas, E. Frazzoli, and K.H. Johansson, ``Distributed
event-triggered control for multi-agent systems," {\em IEEE Trans. Autom. Contr.}, \textbf{57}, 1291--1297, 2012.

\bibitem{Heemels2013}
W.P.M.H. Heemels and M.C.F. Donkers, ``Model-based periodic event-triggered
control for linear systems," {\em Automatica}, \textbf{49}, 698--711, 2013.

\bibitem{Seyboth2013}
 G. S. Seyboth, D. V. Dimarogonas, and K. H. Johansson,
``Event-based broadcasting for multi-agent average consensus," {\em Automatica},\textbf{49}, 245--252, 2013.

\bibitem{Fan2013}
Y. Fan, G. Feng, Y. Wang, and C. Song, ``Distributed event-triggered control for multi-agent systems with combinational measurements," {\em Automatica}, \textbf{49}, 671--675, 2013.

\bibitem{Meng2013}
X. Meng and T. Chen, ``Event based agreement protocols
for multi-agent networks," {\em Automatica}, \textbf{49}, 2125--2132, 2013.

\bibitem{Zhu2014}
W. Zhu, Z. P. Jiang, and G. Feng, ``Event-based consensus of multi-agent systems with general linear models," {\em Automatica}, \textbf{50}, 552--558, 2014.

\bibitem{Garcia2014}
E. Garcia, Y. Cao, and D. W. Casbeer, ``Decentralized event-triggered consensus with general linear dynamics," {\em Automatica}, \textbf{50}, 2633--2640, 2014.

\bibitem{Liuzza2013}
D. Liuzza, D. V. Dimarogonas, M. di Bernardo, and K. H. Johansson,
``Distributed model-based event-triggered control for synchronization
of multi-agent systems," Proc. IFAC Conf. Nonlinear Contr. Syst, 2013,
pp. 329-334.


\bibitem{Adaldo2014}
A. Adaldo, F. Alderisio, D. Liuzza, G. Shi, D. V. Dimarogonas, M. di Bernardo and K. H. Johansson,
``Event-triggered pinning control of complex networks with switching
topologies," Proc. 53rd IEEE CDC, 2014, pp. 2783--2788.



\bibitem{Jadbabaie2003}
A. Jadbabaie, J. Lin, and S. A. Morse, ``Coordination of groups of mobile autonomous agents using
nearest neighbor rules," {\em IEEE Trans. Automat. Contr.}, \textbf{48},
988--1001, 2003.

\bibitem{Ren2007}
W. Ren and E. Atkins, ``Distributed multi-vehicle coordinated control via local
information exchange," {\em Int. J. Robust and Nonlinear Contr.},
\textbf{17}, 1002--1033, 2007.


\bibitem{Hong2006}
Y. G. Hong, J. P. Hu, and L. X. Gao, ``Tracking control for multi-agent
consensus with an active leader and variable topology," {\em Automatica}, \textbf{42}, 1177--1182, 2006.

\bibitem{Hu2011}
J. Hu, G. Chen, and H. Li, ``Distributed event-triggered
tracking control of leader-follower multi-agent systems
with communication delays," {\em Kybernetika}, \textbf{47}, 630--643, 2011.

\bibitem{Zhang2012}
Y. Zhang and Y. Hong, ``Distributed event-triggered tracking control of
multi-agent systems with active leader," Proc. 10th World Congress on Intelligent Control and Automation, Beijing, China, 2012, pp. 1453--1458.

\bibitem{Li2015}
H. Li, X. Liao, T. Huang, and W. Zhu, ``Event-triggering sampling based leader-following
consensus in second-order multi-agent systems," {\em IEEE Trans. Autom. Contr.}, \textbf{60}, 1998--2003, 2015.


\bibitem{Hu2015}
J. Hu, J. Geng and H. Zhu, ``An observer-based consensus tracking control and application to event-triggered tracking", {\em Communications in Nonlinear Science and Numerical Simulation}, \textbf{20}, 559--570, 2015.

\bibitem{Liu2013}
 T. Liu, M. Cao, C. De Persis, and J. M. Hendrickx, ``Distributed
event-triggered control for synchronization of dynamical networks
with estimators", Proc. IFAC Workshop on Distributed Estimation and
Control in Networked Systems, Koblenz, Germany, September 2013,
pp. 116--121.


\bibitem{Franceschelli2013}
M. Franceschelli, A. Gasparri, A. Giua, and C. Seatzu, ``Decentralized
estimation of Laplacian eigenvalues in multi-agent systems," {\em
  Automatica}, \textbf{49}, 1031--1036, 2013.

\bibitem{Zhang2012a}
H. W. Zhang, Frank L. Lewis, and Z. H. Qu, ``Lyapunov, adaptive, and optimal design techniques for cooperative systems on directed communication graphs," {\em IEEE Trans. Industrial Electronics}, \textbf{59}, 3026--3041, 2012.

\bibitem{Hu2007}
J. Hu and Y. Hong, ``Leader-following coordination of multi-agent systems with coupling time delays," {\em Physica A}, \textbf{374}, 853--863, 2007.

%

\bibitem{Kaszkurewicz2000}
 E. Kaszkurewicz and A. Bhaya, Matrix Diagonal in Systems and Computation, {\em Birkh{\"a}user Boston}, 2000.

\bibitem{Zhu2015}
W. Zhu and Z.P. Jiang, ``Event-based leader-following consensus of multi-agent systems with input time delay," {\em IEEE Trans. Automat. Contr.}, \textbf{60},
1362--1367, 2015.

\end{thebibliography}
\end{document}